\documentclass{emulateapj}

\usepackage{graphicx}

\newcommand{\msunyr}{\ensuremath{\mathit{M}_{\odot}{\rm yr}^{-1}}}   
\newcommand{\kms}{\ensuremath{{\rm km\,s^{-1}}}}                   
\newcommand{\msun}{\ensuremath{\mathit{M}_{\odot}}}               
\newcommand{\mini}{\ensuremath{\mathit{M}_\mathrm{ini}}}                
\newcommand{\lsun}{\ensuremath{\mathit{L}_{\odot}}}                  
\newcommand{\rsun}{\ensuremath{\mathit{R}_{\odot}}}                  

\newcommand{\lstar}{\ensuremath{\mathit{L}_{\star}}}                 
\newcommand{\mdot}{\ensuremath{\dot{M}}}                             
\newcommand{\rstar}{\ensuremath{\mathit{R}_{\star}}}                 
\newcommand{\teff}{\ensuremath{\mathit{T}_{\rm eff}}}                
\newcommand{\reff}{\ensuremath{\mathit{R}_{\rm phot}}}                
\newcommand{\vinf}{\ensuremath{v_{\infty}}}                          

\newcommand{\vrot}{\ensuremath{v_{\rm rot}}}                         
\newcommand{\vcrit}{\ensuremath{v_{\rm crit}}}                         
\newcommand{\vsini}{\ensuremath{v \sin{i}}}                         
\newcommand{\ang}{\ensuremath{\mathrm{{\AA}}}}                

\shorttitle{Bona-fide LBVs are fast rotators}
\shortauthors{Groh et al.}

\begin{document}

\title{Bona-fide, strong-variable galactic Luminous Blue Variable stars are fast rotators: detection of a high rotational velocity in HR Carinae\altaffilmark{1}}

\author{J. H. Groh\altaffilmark{2}, A. Damineli\altaffilmark{3}, D. J. Hillier\altaffilmark{4},
R. Barb\'a\altaffilmark{5,7}, E. Fern\'andez-Laj\'us\altaffilmark{6}, R. C. Gamen\altaffilmark{6}, A. Mois\'es\altaffilmark{3}, G. Solivella\altaffilmark{6}, M. Teodoro\altaffilmark{3}}
\email{jgroh@mpifr-bonn.mpg.de}

\altaffiltext{1}{Based on observations made with the 1.6-m telescope at the
Observat\'orio Pico dos Dias (OPD--LNA, Brazil), with the 2.15-m telescope of the Complejo Astronomico El Leoncito (CASLEO, Argentina), and with the 2.2-m ESO telescope at La Silla (Chile) under program 083.D-0589. CASLEO is operated under agreement between CONICET, SECYT, and the National Universities of La Plata, 
C\'ordoba and San Juan, Argentina.}
\altaffiltext{2}{Max-Planck-Institut f\"ur Radioastronomie, Auf dem H\"ugel 69, D-53121 Bonn, Germany}
\altaffiltext{3}{Instituto de Astronomia, Geof\'{\i}sica e Ci\^encias  Atmosf\'ericas, Universidade de S\~ao Paulo, Rua do Mat\~ao 1226, Cidade Universit\'aria, 05508-090, S\~ao Paulo, SP, Brazil}
\altaffiltext{4}{Department of Physics and Astronomy, University of Pittsburgh,
3941 O'Hara Street, Pittsburgh, PA, 15260, USA}
\altaffiltext{5}{Departamento de F\'{\i}sica, Universidad de La Serena, Benavente 980, La Serena, Chile}
\altaffiltext{6}{Facultad de Ciencias Astron\'omicas y Geof\'{\i}sicas, Universidad Nacional de La Plata, and Instituto de Astrof\'{\i}sica de La Plata (CCT La Plata - CONICET), Paseo del Bosque S/N, B1900FWA, La Plata, Argentina }
\altaffiltext{7}{ICATE-CONICET, San Juan, Argentina}
\begin{abstract}

We report optical observations of the Luminous Blue Variable (LBV) HR~Carinae which show that the star
has reached a visual minimum phase in 2009. More importantly, we detected absorptions
due to \ion{Si}{4} $\lambda\lambda$4088--4116. To match their observed line profiles from 2009 May, a high rotational velocity of $\vrot\simeq 150 \pm 20~\kms$ is needed (assuming an inclination angle of $30\degr$), implying that HR Car rotates at $\simeq0.88 \pm 0.2$ of its critical velocity for break-up ($\vcrit$).  
Our results suggest that fast rotation is typical in all strong-variable, bona-fide galactic LBVs, which present S Dor-type variability. Strong-variable LBVs are located in a well-defined region of the HR diagram during visual minimum (the ``LBV minimum instability strip"). We suggest this region corresponds to where $\vcrit$ is reached. To the left of this strip, a forbidden zone with $\vrot/\vcrit>1$ is present, explaining why no LBVs are detected in this zone. Since dormant/ex LBVs like P Cygni and HD 168625 have low $\vrot$, we propose that LBVs can be separated in two groups: fast-rotating, strong-variable stars showing S-Dor cycles (such as AG Car and HR Car) and slow-rotating stars with much less variability (such as P Cygni and HD 168625). We speculate that SN progenitors which had S-Dor cycles before exploding (such as in SN 2001ig, SN 2003bg, and SN 2005gj) could have been fast rotators. We suggest that the potential difficulty of fast-rotating Galactic LBVs to lose angular momentum is an additional evidence that such stars could explode during the LBV phase.
\end{abstract}

\keywords{stars: atmospheres --- stars: mass loss --- stars: variables: other --- supergiants --- stars: individual (HR Carinae) --- stars: rotation}

\section{Introduction}  \label{intro}

Massive stars are the main contributors to the input of ionizing photons, energy, and momentum into the interstellar medium and are responsible for a significant fraction of the chemical enrichment of their host galaxy. Evolutionary models predict the existence of a short-lived, transitional stage, usually referred to as the Luminous Blue Variable (LBV) phase \citep{conti84,hd94}, characterized by a high mass-loss rate ($\mdot \sim 10^{-5}-10^{-3}$ $\msunyr$). In the current picture of stellar evolution, LBVs are rapidly-evolving massive stars in the transitory phase from being an O-type star burning hydrogen in its core to becoming a Wolf-Rayet (WR) type, helium core-burning star \citep{hd94,maeder_araa00,meynet00,meynet03}.

There is strong evidence that very massive stars lose the bulk of their mass through giant outbursts during the LBV phase \citep{so06}, similar to the one that occurred in Eta Carinae in the 1840's. Surprisingly, recent works suggest that some core-collapse supernovae (SNe) have LBV progenitors \citep[e.g.,][]{kv06,smith07,smith08_2006tf,smith08_2006gy,galyam07,trundle08,galyam09}, which not only dramatically enhances the cosmological importance of LBVs but also poses a great challenge to the current paradigm of massive star evolution.

LBVs are rare: only a dozen or so are known in the Galaxy, and roughly only three of them, namely AG Carinae, HR Carinae, and WRA 751, are confirmed strong-variable LBVs \citep{hd94,vg01,clark05}. These stars have been observed for a sufficiently long time to establish a record of strong photometric and spectroscopic variability which shows, undoubtedly, that both strong S Dor-type variability and a circumstellar nebula are present. The spectroscopic and photometric behavior of these strong-variable LBVs are markedly different compared to other stars which are usually regarded as dormant LBVs, such as P Cygni and Eta Carinae \citep{vg01}.

\defcitealias{bh05}{BH05}
\defcitealias{ghd06}{GHD06}
\defcitealias{ghd09}{GHD09}

AG Car, HR Car, and WRA 751 are unstable and present cyclical strong S Dor-type variability characterized by irregular visual magnitude changes on timescales of decades, with a typical amplitude of $\Delta V \simeq1-2$ mag, and corresponding changes in effective temperature and hydrostatic radius \citep{vg01}. During visual minimum, the star is typically hot, while at visual maximum, a cooler effective temperature is obtained \citep[e.g.,][]{vg82}. How the S Dor-type variability relates to the powerful giant eruptions is not clear, although it could be possible that a relatively large amount of stellar mass, which is not ejected from the star, is taking part in the S Dor-type variability (\citealt{ghd09}; hereafter GHD09). This would suggest that the S Dor-type variability is a {\it failed} giant eruption \citepalias{ghd09}.

At least for AG Car, two additional pieces of the puzzle are known: a significant reduction ($\sim50\%$) in the inferred bolometric luminosity from visual minimum to maximum has been determined \citepalias{ghd09}, and a high rotational velocity has been obtained during minimum (\citealt{ghd06}; hereafter GHD06). Whether the same is true for other bona-fide LBVs, such as HR Car, still remains to be seen.

In this Letter we report photometric and spectroscopic observations of the bona-fide LBV HR Car obtained during 2009 (\S~\ref{hrcobs}), showing that the star has entered a visual minimum phase of the S-Dor cycle. More importantly, we detected absorptions related to the high-excitation lines of
\ion{Si}{4}~$\lambda\lambda$4088--4116 in HR Car. The results of our non-LTE spectroscopic modeling using the radiative transfer code CMFGEN are presented in \S~\ref{modres}. They suggest a photospheric nature for these lines, that HR Car has high rotational velocity ($\vrot$), and that $\vrot$ is very high compared to the critical rotational velocity for break-up ($\vcrit$). In \S~\ref{disc} we discuss the importance of such a detection for understanding the LBV phase and the subsequent evolution of very massive stars.

\section{Observations} \label{hrcobs}

\begin{figure}
\resizebox{\hsize}{!}{\includegraphics{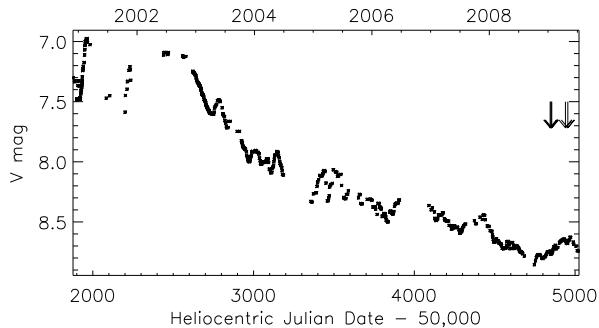}} 
\caption{\label{fig1}ASAS-3 CCD $V$-band lightcurve of HR Carinae from 2001--2009. The arrows indicate epochs when spectroscopic data were available during minimum. }
\end{figure}

Optical CCD $V$-band photometry of HR Car was compiled from the ASAS-3 database \citep{poj02} and averaged in bins of 15 days, with only high-quality (grade A) data being used. The ASAS-3 magnitudes were measured using a large aperture (75\arcsec), but HR Car is by far the dominant source of light in the field. We estimate that the typical error of the 15-day averaged $V$-band magnitude is 0.03 mag. Figure \ref{fig1} presents the $V$-band lightcurve of HR Car from 2001--2009, showing that a significant increase of 1.5 mag occurred in this time span and that HR Car has entered a visual minimum phase of the S-Dor cycle.

Optical spectra of HR Car around the \ion{Si}{4} lines at $\lambda\lambda4088-4116$ were obtained using several facilities in the southern hemisphere. Medium resolution ($R\simeq 12,000$) spectroscopic observations were made at the 1.6-m telescope of the Observat\'orio Pico dos Dias (OPD--LNA, Brazil) and at the 2.15-m telescope of the Complejo Astronomico El Leoncito (CASLEO, Argentina). High-resolution spectra ($R\simeq 48,000$) were gathered using the Fiber-fed Extended Range Optical Spectrograph (FEROS; \citealt{kaufer99}) mounted on the 2.2-m telescope of the Max-Planck-Gesellschaft/European Southern Observatory (MPG/ESO, Chile). The reduction of the spectroscopic data, including bias and sky subtraction, flat-fielding, spectrum extraction, wavelength calibration, correction to the heliocentric frame, and continuum normalization, was carried out using standard IRAF routines. A systemic velocity of $-10~\kms$ \citep{weis97} was subtracted from the observed spectra. Table \ref{obstable} summarizes the observational data of HR Car available during visual minimum. 

\begin{figure}
\resizebox{0.99\hsize}{!}{\includegraphics{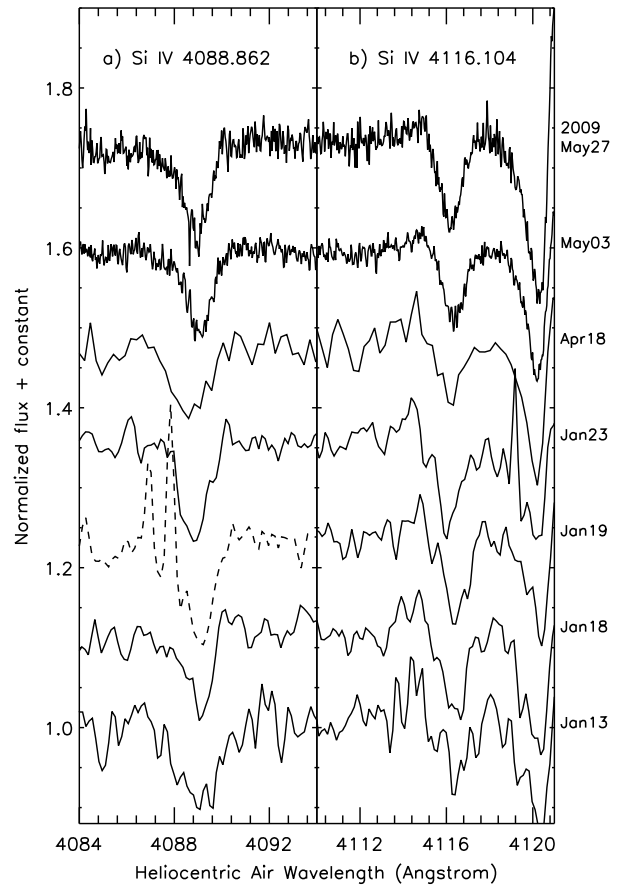}} 
\caption{\label{fig2}Optical spectra of HR Car around \ion{Si}{4} $\lambda\lambda4088-4116$. From bottom to top, we show spectra obtained from 2009 January 13 until 2009 May 27. The blue wing of \ion{Si}{4} $\lambda4088$ was hit by cosmic rays on 2009 January 18th and is shown by the dashed line. }
\end{figure}

Figure \ref{fig2} presents a montage of the observed \ion{Si}{4} $\lambda\lambda4088-4116$ line profiles of HR Car during our 2009 monitoring. The use of different instruments over 5 months confirms that the detection is not transient and, undoubtedly, absorptions due to \ion{Si}{4} $\lambda\lambda4088-4116$ lines are typical in HR Car during visual minimum. 

\begin{deluxetable*}{lcccr}
\tabletypesize{\scriptsize}
\tablecaption{Journal of optical spectroscopy of HR Car during the 2009 visual minimum \label{obstable}}
\tablewidth{0pt}
\tablehead{ \colhead{Date} & \colhead{$V$} & \colhead{Telescope/Instrument} & \colhead{Spectral Range ($\ang$)} & \colhead{$R$} }
\startdata
2009 January 13 & 8.74 & 2.15-m CASLEO/REOSC & 3650--6130 & 12,000 \\
2009 January 18--19--23 & 8.77 & 2.15-m CASLEO/REOSC & 3650--6130 & 12,000 \\
2009 April 18 & 8.66 & 1.6-m OPD--LNA/Coude & 3880--4390 & 12,000 \\
2009 May 03--27 & 8.67 & 2.2-m MPG--ESO/FEROS & 3600--9200 & 48,000 \\
\enddata
\end{deluxetable*} 

\section{Modeling and results } \label{modres}

In the case of the prototypical LBV AG Car, the relatively broad absorptions due to \ion{Si}{4} $\lambda\lambda4088-4116$ are caused by the extremely fast rotation, $\vrot/\vcrit\gtrsim0.86$ \citepalias{ghd06}. Here we also use these \ion{Si}{4} lines to derive the rotational velocity of HR Car, performing a detailed spectroscopic analysis similar to \citetalias{ghd06}, using the radiative transfer code CMFGEN \citep{hm98} and the two-dimensional (2-D) code of \citet[hereafter BH05]{bh05}.

CMFGEN solves the statistical equilibrium equations simultaneously to the radiative transfer in the co-moving frame of the outflow, in spherical symmetry, and assuming steady state. The code computes line and continuum formation in the non-LTE regime, and each model is specified by the stellar luminosity $\lstar$ and effective temperature $\teff$\footnote{The effective temperature is defined as the temperature where the Rosseland optical depth is 2/3, i.e., $\teff=T(\tau_\mathrm{Ross}=2/3$).}, the wind mass-loss rate \mdot, volume filling factor $f$, and terminal velocity \vinf, and the chemical abundances $Z_i$ of the included species. Since the momentum equation of the wind is not solved, a beta-type velocity law is adopted and modified at depth to smoothly match a hydrostatic structure at $\rstar$. The hydrostatic equation is iterated in order to produce a quasi-hydrostatic structure which extends inwards until the inner boundary of the model (at a Rosseland optical depth of 100). Line blanketing is fully taken into account, and our models include tens of thousands of spectral lines in non-LTE. The atomic model used for HR Car is similar to the one used for AG Car \citepalias{ghd09} and included lines of H, He, C, N, O, Na, Mg, Al, Si, Cr, Mn, Fe, Co, and Ni. 

We assumed a depth-independent turbulent velocity of 20\,km\,s$^{-1}$ in the computation of the observed spectrum, which is an upper limit, since a higher turbulent velocity would yield a detectable redshift, which is not observed, of the emission lines with high opacity, such as the stronger lines of \ion{H}{1} and \ion{He}{1}. Therefore, microturbulence cannot explain the broadening seen in the high-ionization \ion{Si}{4} $\lambda\lambda$ 4088--4116 lines, and we suggest these lines to be rotationally broadened (note that other broadening mechanisms, such as convection or other kind of surface activity, have yet to be detected in LBVs and, thus, are not expected to be present in AG Car, although our data do not rule them out).

The effects of rotation on the line profiles were investigated using the \citetalias{bh05} code, which allows the computation of the spectrum in 2-D geometry for a given inclination angle $i$. No equatorial or polar density enhancements were taken into account, since our goal was to investigate the broadening of the photospheric lines. Therefore, we computed a grid of synthetic line profiles with a stellar rotational velocity ranging from 0 to $200~\kms$ in increments of 5~\kms, and a best fit was obtained by visual inspection. We assume that the stellar rotational axis of HR Car is aligned with the polar axis of its bipolar nebula, which is inclined at $i=30\degr \pm10\degr$ \citep{nota97}. 

In a dense stellar wind, ionization stratification occurs, and lines from ions of different ionization potentials will form at different distances $r$ from the star. The \ion{Si}{4} $\lambda\lambda4088-4116$ lines are the highest ionization lines present in the optical spectrum of HR Car and, therefore, should be formed closest to the stellar photosphere. Since the azimuthal velocity component is proportional to $r^{-1}$, \ion{Si}{4} $\lambda\lambda4088-4116$ are ideal for deriving the rotational velocity of HR Car. Our modeling shows that other optical lines which are ocasionally seen in absorption in LBV spectra, such as \ion{N}{2} $\lambda\lambda4601-4607-4630-4643$ and \ion{Si}{3} $\lambda\lambda4553-4558-4575$, are actually formed in the wind of HR Car, far from the stellar photosphere. As a consequence, their broadening is dominated by the wind velocity field (see discussion in \citetalias{bh05} and \citetalias{ghd06}).

\begin{figure}
\resizebox{\hsize}{!}{\includegraphics{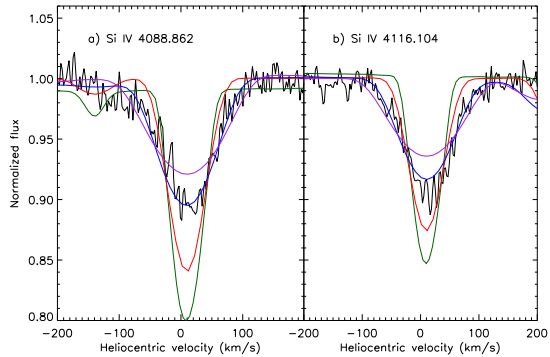}} 
\caption{\label{fig3}Comparison between the \ion{Si}{4} $\lambda\lambda$ 4088--4116 line profiles of HR Car observed on 2009 May 3rd with synthetic model spectra computed with the 2-D code from \citetalias{bh05} for different rotational velocities, assuming $i=30\degr$: $0~\kms$ (green), $75~\kms$ (red), $150~\kms$ (blue), and $200~\kms$ (purple). }
\end{figure}

Figure \ref{fig3} compares the absorption profiles of \ion{Si}{4} $\lambda\lambda$ 4088--4116 observed in 2009 May 03 (which has the highest signal-to-noise ratio and spectral resolution of our sample) with our best CMFGEN model, which has the following parameters: $\lstar=5.0\times10^{5}~\lsun$ (assuming A$_V=3.1$ and $d=5$~kpc, \citealt{vg91,huts91}), $\rstar=70~\rsun$, $\reff=75~\rsun$, $\teff=17900~\mathrm{K}$, He/H=$0.4$ (by number), $\vinf=120~\kms$, $\beta=1$, $\mdot=7\times10^{-6}~\msunyr$, and $f=0.25$. In a forthcoming full paper we will present the results of the detailed spectroscopic analysis of HR~Car during minimum, including diagnostics for each of the parameters assumed above. Nevertheless, for reasonable changes in the above parameters, the derived rotational velocity of HR~Car changes only slightly and the conclusions of this Letter remain valid.

Similar to AG Car \citepalias{ghd06}, the observed absorption profiles of both \ion{Si}{4} lines are significantly broader and shallower than the non-rotating model. Our model predicts that the \ion{Si}{4} lines are formed very close to the stellar photosphere, where the wind is still accelerating and, therefore, the line broadening caused by the wind velocity field is negligible. Consequently, most of the line broadening should be due to rotation. According to our rotating models computed with the \citetalias{bh05} code, a rotational velocity of $\vrot\simeq 150 \pm 20~\kms$ is required to reproduce the observed shape of the \ion{Si}{4} $\lambda\lambda$4088--4116 lines in HR Car (Fig. \ref{fig3}).

How does the rotational velocity of HR Car compare to its critical velocity for break-up? An estimation of $\vcrit$ can be calculated following \citet{mm_omega00} once the stellar mass $M$, \lstar, \rstar, and the Eddington parameter $\Gamma$ are known. The critical parameter for obtaining a precise value of $\vrot/\vcrit$ is $M$, which can be estimated based on the evolutionary models from \citet{meynet03}. Such models predict that, based on our derived value of $\lstar=5.0\times10^{5}~\lsun$, the initial mass of HR Car should have been roughly $\mini \simeq 50 \pm 10~\msun$. While there are no evolutionary models available in the literature specifically for $\mini=50~\msun$, we estimate that stars such as HR Car will reach the LBV phase with $M \sim 25~\msun$, since evolutionary models predict that a $40~\msun$ star reaches the LBV phase with 20~\msun, while a $60~\msun$ star reaches the LBV phase with $30~\msun$ \citep{meynet03}.

Assuming an admittedly uncertain $M=25~\msun$, our CMFGEN model of HR Car predicts $\Gamma\simeq 0.80$, and we derive $\vcrit\simeq 170 \pm 20~\kms$ and $\vrot/\vcrit\simeq0.88 \pm 0.2$. Therefore, similar to AG Car \citepalias{ghd06}, HR Car is also likely rotating very close to, if not {\it at}, its critical velocity for break-up.

\section{Discussion: strong-active, bona-fide galactic LBVs are fast rotators } \label{disc}

Several theoretical works have long predicted that LBVs should have high rotational velocities \citep[e.g.][]{hd94,langer99,mm_omega00,md01,do02,alm04}, but direct observational detection through the broadening of spectral lines has been elusive until recently, in particular because high-resolution observations of LBVs in the blue part of the optical spectrum are scarcely available during visual minimum. The detection of high rotational velocity in two of the most variable LBVs in the Galaxy, AG Car \citepalias{ghd06} and now HR Car (\S~\ref{modres}), raises the obvious question: {\it are all LBVs fast rotators?}

The presence of fast rotation in two out of three\footnote{Unfortunately, no high-resolution observations of WRA 751 during minimum are available to us to check if it is a fast rotator.}  of the galactic bona-fide LBVs that have a circumstellar nebula and strong S Dor-type variability is probably not a coincidence, and {\it we suggest that fast rotation is typical in such stars.} Perhaps, fast rotation might even be a key ingredient for the instability that causes the S-Dor cycles in LBVs. If that is true, we should expect that fast rotation is also present during visual minimum in extragalactic LBVs, such as R127 and S Doradus in the LMC. An obvious test case for such a hypothesis would be R127, which has been undergoing a visual minimum phase since 2007 \citep{walborn08}.

\begin{figure}
\resizebox{\hsize}{!}{\includegraphics{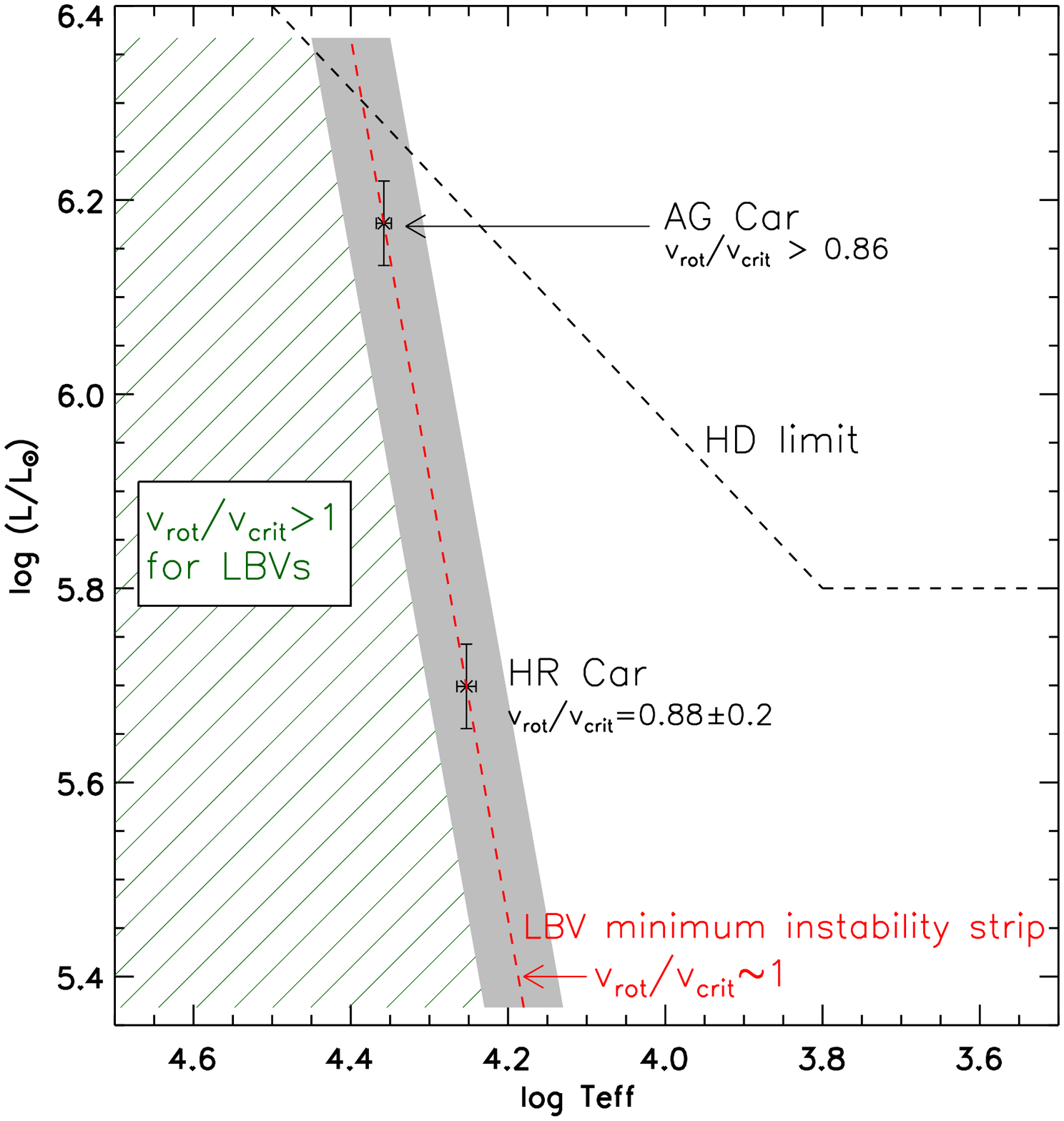}} 
\caption{\label{fig4}HR diagram showing the position of the strong-variable LBVs AG Car and HR Car during visual minimum, according to the updated stellar parameters determined in \citetalias{ghd09} and in this work. A revised position for the LBV minimum instability strip is provided (red dashed line), and is considerably steeper than previous determinations \citep{wolf89,vg01,clark05}. Notice that AG Car and HR Car are both fast rotators during minimum, which suggests that the LBV minimum instability strip corresponds to the location where critical rotation is reached for strong-variable LBVs. The location of the Humphreys-Davidson limit \citep{hd94} is shown (black dashed line).}
\end{figure}

Figure \ref{fig4} presents the position of AG Car and HR Car in the HR diagram based on their updated stellar parameters obtained in \citetalias{ghd09} and in this work. Previous works have recognized that, during visual minimum, LBVs are located in a well-defined region of the HR diagram, the so-called ``LBV minimum instability strip" \citep{wolf89,clark05}. Our results suggest that the LBV minimum instability strip is considerably steeper than previously determined, being characterized by $\log{(L/\lsun)}=4.54\,\log{(\teff/\mathrm{K})}$ - 13.61 (red dashed line in Fig. \ref{fig4}). {\it More importantly, we suggest that the LBV minimum instability strip corresponds to the region where critical rotation is reached for LBVs with strong S-Dor type variability.} When LBVs are evolving towards maximum, the star moves far from the LBV minimum instability strip (to the right in the HR diagram), and $\vrot/\vcrit$ decreases considerably \citepalias{ghd06}. The region in the HR diagram on the left side of the LBV minimum instability strip would be populated by unstable LBV stars with $\vrot/\vcrit > 1$, which would make it a ``forbidden region" for LBVs. Indeed, no confirmed strong-variable LBV is seen in this region \citep{vg01,clark05}.

A corollary of the relationship between S Dor-type instability and fast rotation that we suggest in this Letter regards SNe which likely had an LBV progenitor. Some of these LBVs showed evidence for S Dor-type variability before exploding as a SN, such as the progenitors of SN 2001ig, SN 2003bg \citep{kv06}, and SN 2005gj \citep{trundle08}. It could be very likely that these massive stars were fast rotators before exploding.

What about dormant LBVs which have not presented evidence for strong S Dor-type variability for a century or more? From these, only Eta Car has shown indirect evidence of fast rotation, through the presence of a latitude-dependent wind \citep{smith03,weigelt07}. Interestingly, at least two other confirmed LBVs that do not show S Dor-type variability appear to rotate slowly. P Cygni has evidence of a relatively low rotational velocity of $50-60~\kms$ \citep{najarro97}\footnote{Assuming the rotational axis of P Cygni is aligned with part of its nebula, which is seen at $i\sim45\degr$ \citep{smith06b}. Note that this assumption is not crucial for the results of this paper.}, a property also shared by HD 168625, for which we measured $\vsini \simeq 40~\kms$ based on optical photospheric \ion{N}{2} lines and a similar methodology as in the present work (J. H. Groh 2009, in preparation). Assuming $i\simeq60\degr$ \citep{smith_hd168625}, we obtain $\vrot\simeq 45 \kms$. For both P Cygni and HD 168625, the critical velocity for break-up should be of the order of 150--200~\kms, meaning that $\vrot/\vcrit$ should be $\sim0.2$--0.3 for these stars.

We suggest that LBVs, at least in our Galaxy, can be separated into two groups: group I, comprised of fast-rotating, strong-variable LBVs such as AG Car and HR Car, and group II, composed of slow-rotating, weak variable LBVs such as P Cygni and HD 168625. While an LBV can in principle migrate from one subgroup to the other, for instance after a giant eruption, these LBV groups might also follow different evolutionary paths during and after the LBV phase, possibly yielding significantly different SN progenitors. 

In particular, fast-rotating LBVs might struggle to lose enough angular momentum through winds and giant eruptions. Fast-rotating stars are supposed to have a latitude-dependent, polar-enhanced wind \citep{owocki96} which, depending on the mass-loss rate, will not remove as much angular momentum as in the case of a spherical wind (see discussion in \citealt{maeder02}). In addition, LBVs might not lose enough angular momentum during giant outbursts, similar to the one that occurred in Eta Car in the 1840's, since almost 75\% of the mass was lost at latitudes higher than $45\degr$ \citep{smith06}. Although current evolutionary models predict that fast-rotating LBVs should spin-down \citep{meynet03}, the models do not include angular momentum and mass losses through a giant outburst, which is a key factor during the LBV phase \citep{so06}. In the Galaxy, there are at least three examples of stars which had a giant eruption, lost significantly amounts of mass (several \msun), but are still currently fast rotators: Eta Car \citep{smith03,weigelt07}, AG Car \citep{ghd06}, and HR Car (this work).
 
As a consequence, if fast-rotating LBVs are to evolve to WR stars without losing enough angular momentum, they would produce a population of significantly fast-rotating WR stars, which is yet unseen in the Galaxy\footnote{\citet{massey80} suggested a high rotational velocity of $\sim500~\kms$ for the WN5 star HD 193077, based on broad absorptions due to high Balmer and \ion{He}{1} lines. However, \citet{annuk90} proposed that HD193077 is a WR+O binary and that the broad absorption lines are formed in the O star. Even if the absorption lines were formed in the WR wind, they should be formed sufficiently far from the hydrostatic core that the broadening would be dominated by the wind acceleration.} and not predicted by evolutionary models \citep{meynet03}. An alternative evolutionary path would be that LBVs could fail to lose an additional significant amount of mass and thus, instead of becoming a Wolf-Rayet star, would dramatically explode as a core-collapse supernova anytime during the LBV phase \citep[e.g,][]{smith07,smith08}.
%
We suggest that the potential difficulty of fast-rotating Galactic LBVs to lose angular momentum is an additional evidence that LBVs such as AG Car and HR Car could explode during the LBV phase.

\acknowledgments

We thank Nathan Smith, Gerd Weigelt, Thomas Driebe and an anonymous referee for invaluable comments. JHG thanks the Max Planck Society for financial support. AD, AM, and  MT acknowledge support from FAPESP and CNPq, DJH from NSF grant AST-0507328, and RHB partially from Universidad de La Serena (project DIULS CD08102).


{\it Facilities:} \facility{OPD/LNA}, \facility{CASLEO}, \facility{ESO}

\end{document}